\documentclass[letterpaper, 10 pt, conference]{ieeeconf}  % Comment this line out if you need a4paper

\IEEEoverridecommandlockouts                              % This command is only needed if 
\usepackage{graphics} % for pdf, bitmapped graphics files
\usepackage{epsfig} % for postscript graphics files
\usepackage{times} % assumes new font selection scheme installed
\usepackage{amsmath} % assumes amsmath package installed
\usepackage{amssymb}  % assumes amsmath package installed
\usepackage{float}
\usepackage{graphicx}
\usepackage{subfigure}
\usepackage{tcolorbox}
\usepackage{cuted}
\usepackage{mathtools}
\usepackage{url}
\usepackage{setspace}
\usepackage{dblfloatfix}

\title{Multi-Agent Phase-Balancing around Polar Curves with Bounded Trajectories: An Experimental Study using Crazyflies and MoCap System}

\author{Gaurav Singh Bhati, KKN Shyam Sathvik, Anuj Patil and  Anoop Jain% <-this % stops a space
\thanks{All authors are with the Department of Electrical Engineering, Indian Institute of Technology Jodhpur, 342030, India. The support from the MeitY project (S/MeitY/SKS/20210139) is gratefully acknowledged. %Gaurav contributed throughout the setup of the experiment both hardware and software and did all the experiments. Shyam contributed throughout the setup of ROS2 and modifying the crazyflie\(\_\)server. Anuj contributed throughout the setup of ROS2 and hardware implementation. The support from the MeitY project (S/MeitY/SKS/20210139) is gratefully acknowledged.
}
}

\begin{document}

\maketitle
\thispagestyle{empty}
\pagestyle{empty}

%%%%%%%%%%%%%%%%%%%%%%%%%%%%%%%%%%%%%%%%%%%%%%%%%%%%%%%%%%%%%%%%%%%%%%%%%%%%%%%%
\begin{abstract}
In this experimental work, we implement the control design from existing work \cite{hegde2023synchronization} on a swarm of Crazyflie 2.1 quad-copters by deriving the original control in terms of variables that are available to the user in this practical system. A suitable model is developed using the Crazyswarm2 package within ROS2 to facilitate the execution of the control law. We also discuss various components that are part of this experiment and the challenges we encountered during the experimentation. Extensive experimental results, along with the links to the YouTube videos for actual Crazyflie quad-copters, are provided.  
\end{abstract}

%%%%%%%%%%%%%%%%%%%%%%%%%%%%%%%%%%%%%%%%%%%%%%%%%%%%%%%%%%%%%%%%%%%%%%%%%%%%%%%%
\section{Introduction}
\subsection{Motivation}
Multi-agent formation control such that the agents' trajectories remain confined within an external boundary has garnered significant attention, particularly in the context of safe border surveillance missions. One recent contribution in this domain can be found in \cite{hegde2023synchronization}, where the objective is to stabilize unicycle agents in a phase-balanced configuration along simple closed polar curves, ensuring that their trajectories remain bounded within a compact set during stabilization. In phase-balancing configuration, the agents are equally dispersed on the desired closed curve. Although a good amount of theoretical work has been done recently in this direction exploiting the ideas of control barrier functions and barrier Lyapunov functions, the practical implementations remain relatively unexplored. This paper aims to implement the control design of \cite{hegde2023synchronization} on Crazyflie 2.1 quad-copters in a motion capture system (MoCap) environment. 
\subsection{Objectives and Contributions}
%The main features of this work are as follows: 
\begin{enumerate}
	\item[(a)] We practically implement the control law for phase-balancing of three Crazyflie 2.1 quadcopters about circular and elliptical paths. It is illustrated that the quad-copters are equally spaced around the desired circular/elliptical orbit and their trajectories remain bounded during stabilization as in \cite{hegde2023synchronization}. 
	\item[(b)] We prepare the experimental setup for controlling the Crazyflie 2.1 quad-copter as a unicycle model by modifying the \emph{crazyflie\_server} and relevant back-end files in the Crazyswarm2 application programming interface. We have used the MoCap system for the purpose of getting feedback on pose data for the deployed quad-copters. 
\end{enumerate}
\subsection{Paper Structure and Notations}
We begin with summarizing the key results of \cite{hegde2023synchronization} in Section~\ref{section_1}, where we also provide simplification of the control law for elliptical and circular trajectories and a few simulation results validating it. The experimental setup is described in Section~\ref{section_2}, followed by experimental results in Section~\ref{section_3}. A brief discussion on the challenges and the future aspects are discussed in Section~\ref{section_4}. Throughout the paper, $\mathbb{R}$, $\mathbb{R}_{>0}$ and $\mathbb{C}$ denotes the set of real, positive-real and complex numbers, respectively. For $z \in \mathbb{R}^n$, $z^T$ is its transpose. 

\section{Summary of \cite{hegde2023synchronization} and Simulation Results}\label{section_1}
In this work, we model Crazyflie quad-copters as unicycle robots operating at a constant height. This permits us to operate these in a planar space with the following kinematics  
\begin{equation}\label{unicycle}
	\dot{x}_k  = v_k\cos \theta_k, \ \dot{y}_k = v_k \sin \theta_k, ~ \dot{\theta}_k = u_k,
\end{equation}
for $ k = 1, \ldots, N$. Here, $[x_k, y_k]^T \in \mathbb{R}^2$ is the position, $\theta_k \in [0, 2\pi)$ is the heading angle, and $u_k \in \mathbb{R}$ is the turn rate controller for the $k$-th agent. In the case of the Crazyflie robot, the heading angle is equivalent to the yaw angle. For simplicity, we consider that the linear speed $v_k = 1, \forall k$, and represent the model \eqref{unicycle} in the complex plane as 
\begin{equation}\label{unicycle_complex_plane}
	\dot{r}_k  = {\rm e}^{i\theta_k}, ~ \dot{\theta}_k = u_k, \ k = 1, \ldots, N,
\end{equation}
where $r_k = x_k + iy_k \in \mathbb{C}$ is the position of the $k$-th agent. 

\begin{figure}[t!] 	
	\centering
	\hspace*{-0.5cm}	\subfigure[Circular curve, $r = 1$]{\includegraphics[width=4.0cm]{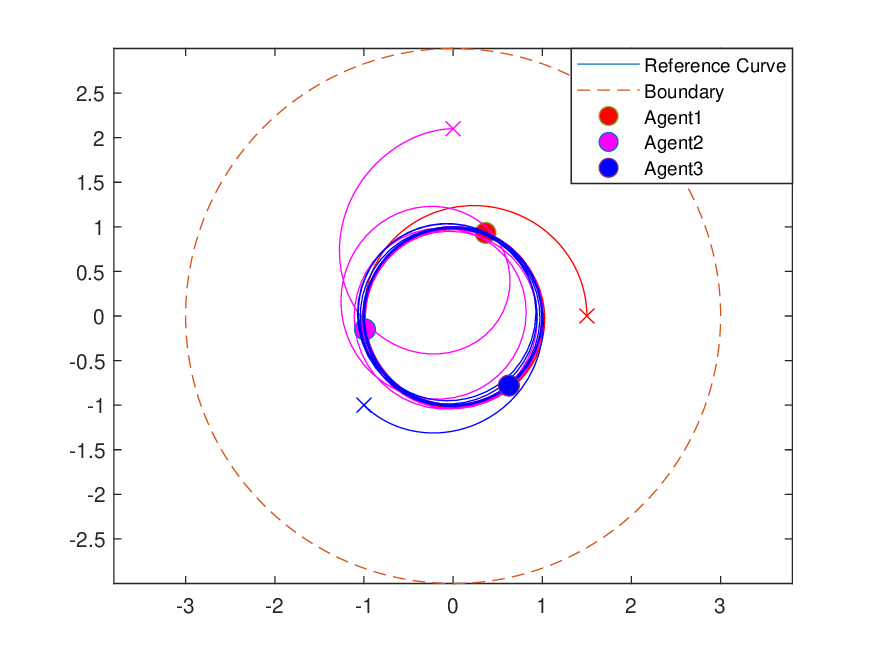}\label{circle}}\hspace*{-0.25cm}
	\subfigure[Elliptical curve, $a = 2, b = 1$]{\includegraphics[width=4.0cm]{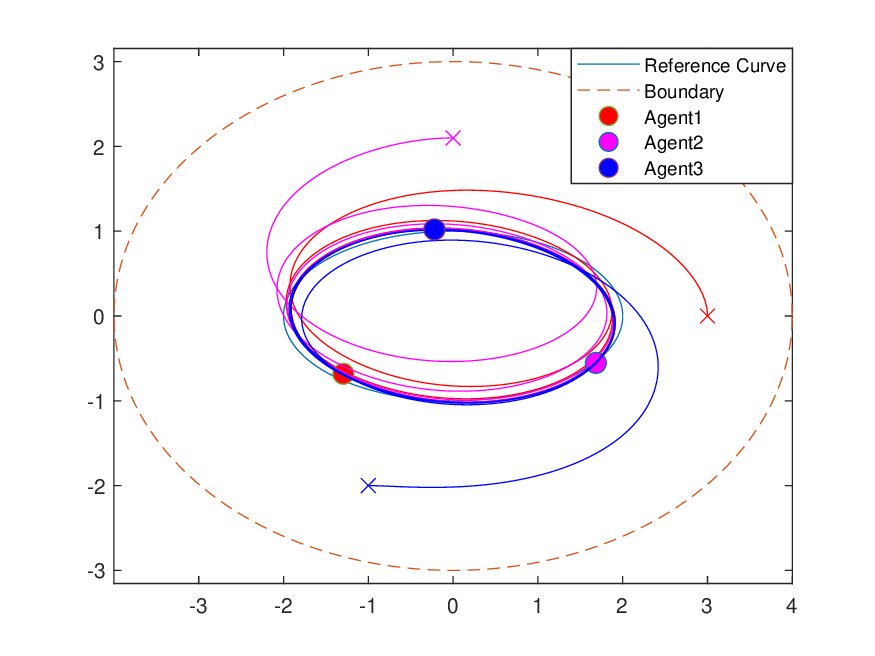}\label{ellipse}} 	 
	\caption{Trajectory-constrained balancing: circular and elliptical orbits.}
	\label{simulations}
	\vspace*{-15pt}
\end{figure}

\begin{figure*}[t!]
	\centering
	\hspace*{-15pt}\subfigure[Gate view]{\includegraphics[scale=0.136]{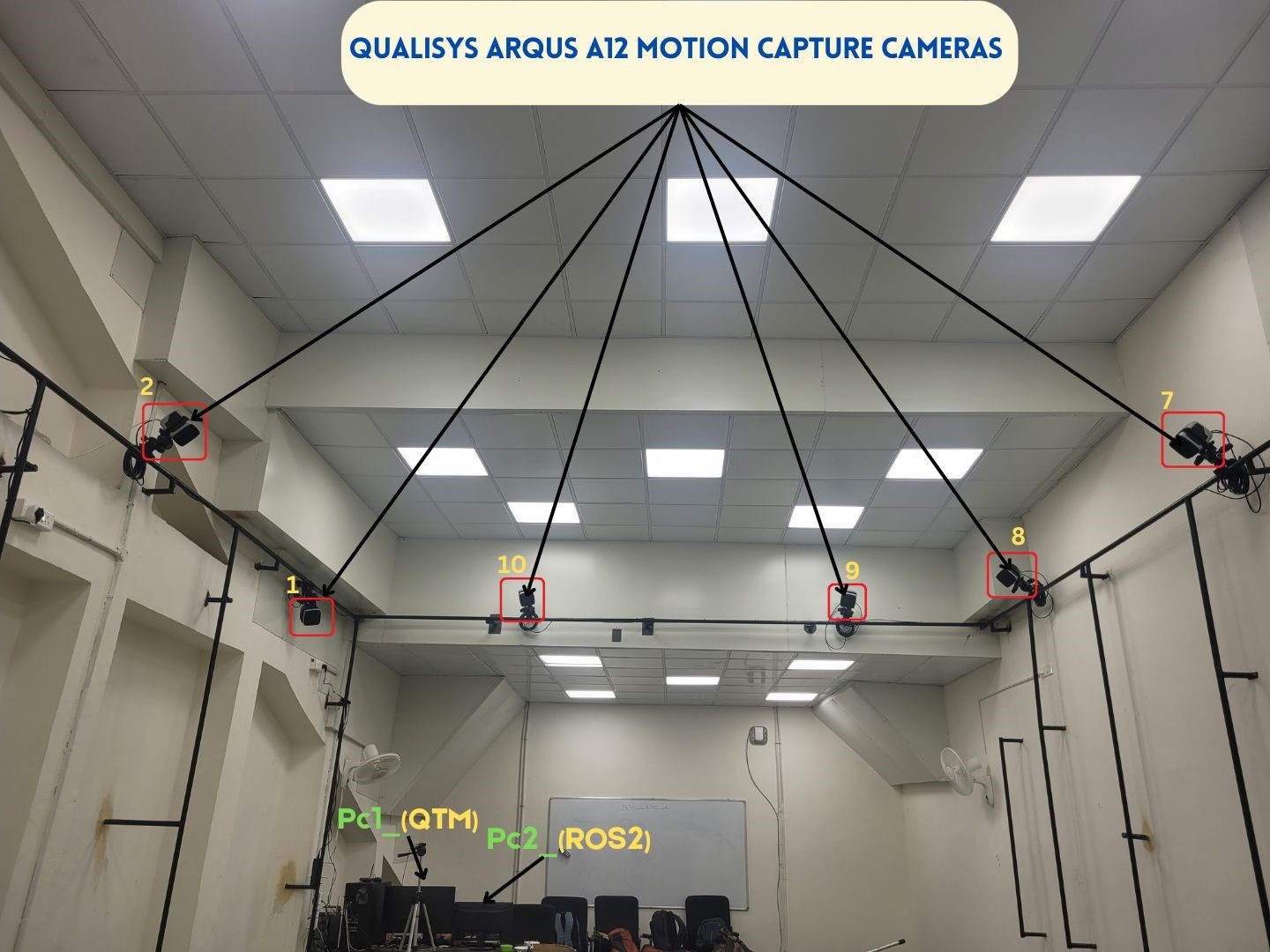}\label{mocap1}} \hspace*{0.1cm}
	\subfigure[Desk view]{\includegraphics[scale=0.136]{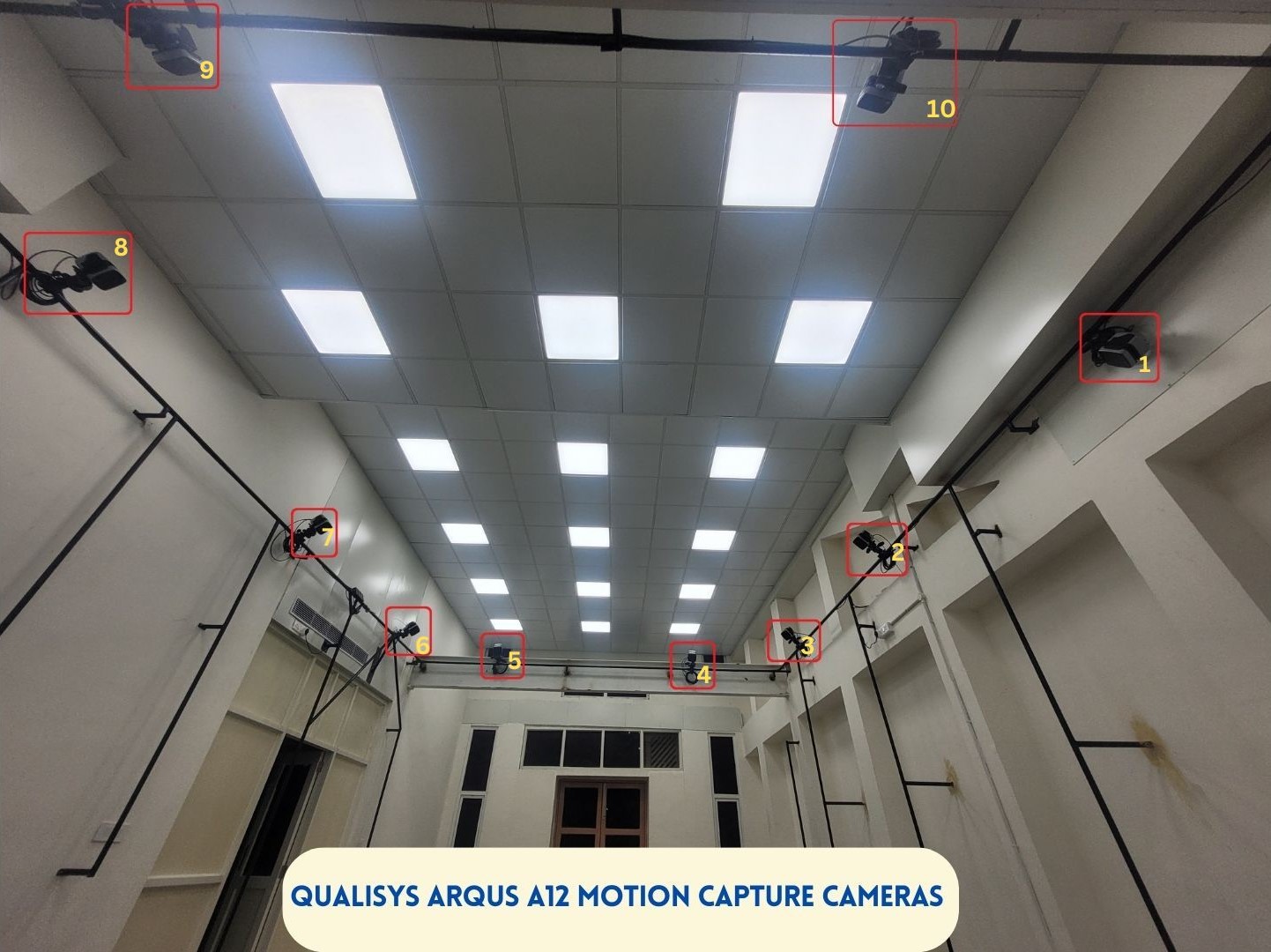}\label{mocap2}} \hspace*{0.1cm}
	\subfigure[Calibrated volume]{\includegraphics[scale=0.196]{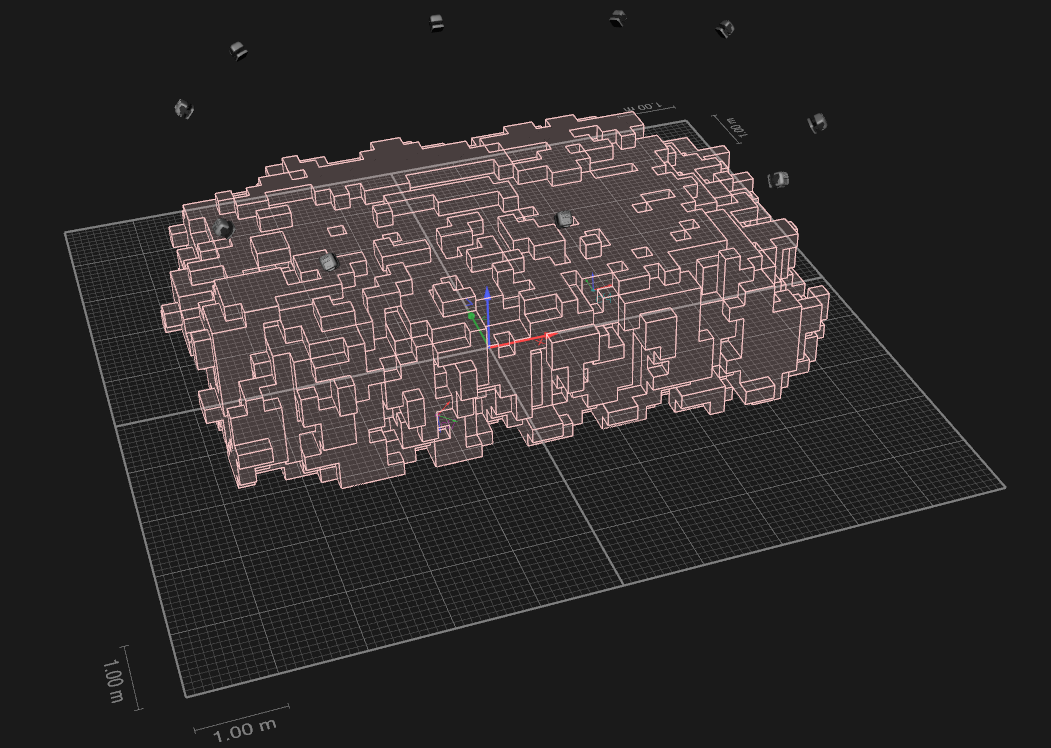}\label{3dvolume}} 
	\caption{The MoCap System setup.}	
	\label{mocap}
	\vspace*{-12pt}
\end{figure*}

\subsection{Control Law}
Considering model \eqref{unicycle_complex_plane}, \cite{hegde2023synchronization} proposed control $u_k$, by leveraging the concept of barrier Lyapunov function, for stabilizing agents motion around a family of polar curves, parametrized by $\phi$ with respect to the origin, and expressed in polar representation as $\rho(\phi) = R(\phi){\rm e}^{i\phi} = R(\phi)\cos\phi + i R(\phi)\sin\phi \in \mathbb{C}, \ R(\phi) \in \mathbb{R}_{>0}$. Let $e_k = r_k - \rho$ be the positional error of the $k$-th agent with respect to the desired curve. The control law for the $k$-th agent is proposed as \cite{hegde2023synchronization}:
\begin{equation} \label{control}
	u_k = \kappa(\phi)(1 + \zeta_k),
\end{equation}
where $\kappa(\phi)$ is the curvature of the desired curve and $\zeta_k$ is derived from the formation control objectives of the group and is given by
\begin{align}\label{zeta}
	\nonumber \zeta_k = & K_c \left[\frac{(x_k  - R(\phi)\cos\phi) \cos\theta_k + (y_k - R(\phi)\sin\phi)\sin \theta_k}{\delta^2- |e_k|^2}\right] \\
	& ~~~~~~~~~~~~~~~~~~~~~~~~ -\frac{K}{N}\sum_{j=1}^{N} \sin(\psi_j-\psi_k),
\end{align}
where $K_c > 0$ and $K > 0$ are control gains, and $\delta > 0$ is the uniform radial distance of the external boundary from the desired curve. Further, $\psi_k$, defined as
\begin{equation} \label{psi_k}
\psi_k = (2\pi/\Gamma)\sigma_k,
\end{equation}
is the curve-phase of the agent with respect to the desired curve, with perimeter $\Gamma$ and arc-length $\sigma_k$ (please refer to \cite{hegde2023synchronization} for further details). Please note that, in \eqref{zeta}, the first term is responsible for bringing all the agents on the desired curve while restricting their trajectories, and the second term allows the agents to have the phase-balancing formation on the desired curve. It is shown in \cite{hegde2023synchronization} that $\zeta_k \to 0, \forall k$ at the steady-state, implying that the agents move at the desired curve in the phase-balancing. Further, the agents' trajectories remain bounded within a compact set characterized by the radial distance $\delta$. The proposed controller \eqref{control} is applicable to all initial conditions such that $|e_k(0)| < \delta$ for all $k$.

In this experimental work, we focus on achieving such trajectory-constrained formations about two closed orbits, namely circular orbit and elliptical orbit. Below, we discuss how to control \eqref{control} shape for both these cases: 
\subsubsection{Elliptical Orbit}
In the case of an ellipse, the coordinate of any point on the curve can be expressed in terms of the parametrization $\phi$ as $\rho(\phi) = a\cos\phi + i b\sin\phi$, where $a$ and $b$ are the major and minor axis of the ellipse, respectively. Further, one can easily derive the relation between $\phi$ and the heading angle $\theta_k$, which is given by
\begin{equation}\label{angle_relation}
	\tan\phi =  -({b}/{a})\cot \theta_k,
\end{equation}
using which, the radial distance $R(\phi)$ and the curvature $\kappa(\phi)$ can be expressed in terms of the heading angle $\theta_k$ (which is essentially available to us in the Crazyflie robot) as \cite{1123737}:
\begin{align}
	\label{r_theta} R(\theta_k) &= \sqrt{a^2\sin^2\theta_k + b^2\cos^2\theta_k}\\
	\label{kappa_theta} \kappa(\theta_k) &= \frac{\sqrt{(a^2\sin^2\theta_k + b^2\cos^2\theta_k)^3}}{a^2 b^2}. 
\end{align}
As defined in \eqref{psi_k}, the perimeter and arc length of the elliptical curve are given by \cite{1123737}
\begin{align}
	\Gamma &=  \pi\left[3(a+b)-\sqrt{(3a+b)(a+3b)}\right] \\
	\sigma_k &= b\mathbb{E}\left[\arctan\left(\frac{a}{b})\tan\theta_k\right) \ {\Big |} \ 1- \left(\frac{a}{b}\right)^2\right],
\end{align}
where $\mathbb{E}(\bullet \ | \ \star)$ is an incomplete elliptical integral of the second type \cite{1123737}.

\begin{figure}[t!]
	\centerline{\includegraphics[scale=0.5]{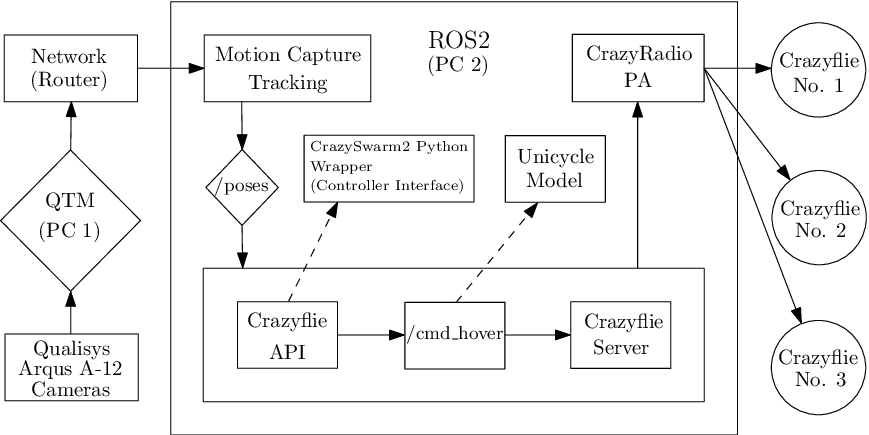}}
	\caption{Block diagram of data flow in experimental setup.}
	\label{rqt}
	\vspace*{-3pt}
\end{figure} 

 \begin{figure}[t!]
	\centering
	\includegraphics[scale=0.05]{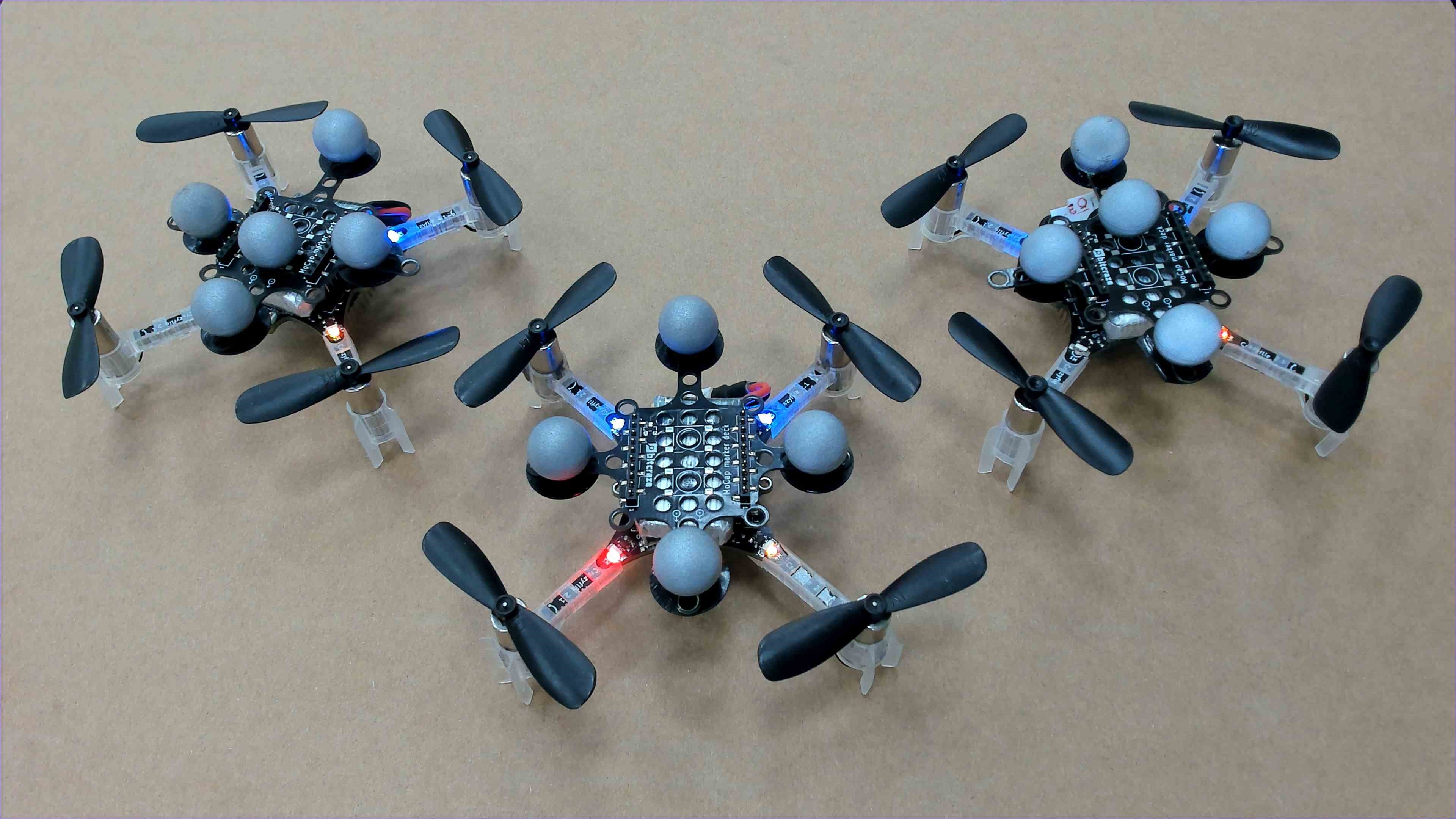} 
	\caption{Swarm of 3 Crazyflie 2.1 quadcopters with passive marker deck.}
	\label{cf_three}
	\vspace*{-15pt}
\end{figure}

\subsubsection{Circular Orbit}
In case of circular orbit, $a = b = r$, and hence, $R = r$ and $\kappa = 1/r$ are constant, using \eqref{r_theta} and \eqref{kappa_theta}, respectively. Further, $\phi = (\pi/2 + \theta_k)$ using \eqref{angle_relation}, and $\psi_k = \theta_k$, using \eqref{psi_k}, as $\Gamma = 2\pi r$. Consequently, the control law \eqref{zeta} can be written in the simplified form as:
 \begin{equation}\label{zeta_circle}
\hspace*{-15pt}\zeta_k = K_c \left[\frac{x_k \cos\theta_k + y_k \sin \theta_k}{\delta^2- |e_k|^2}\right] - \frac{K}{N}\sum_{j=1}^{N} \sin(\theta_j-\theta_k).
 \end{equation} 
 
 \begin{figure*}[t]
 	\centering
 	\subfigure[Take-off from random positions]{\includegraphics[scale=0.0483]{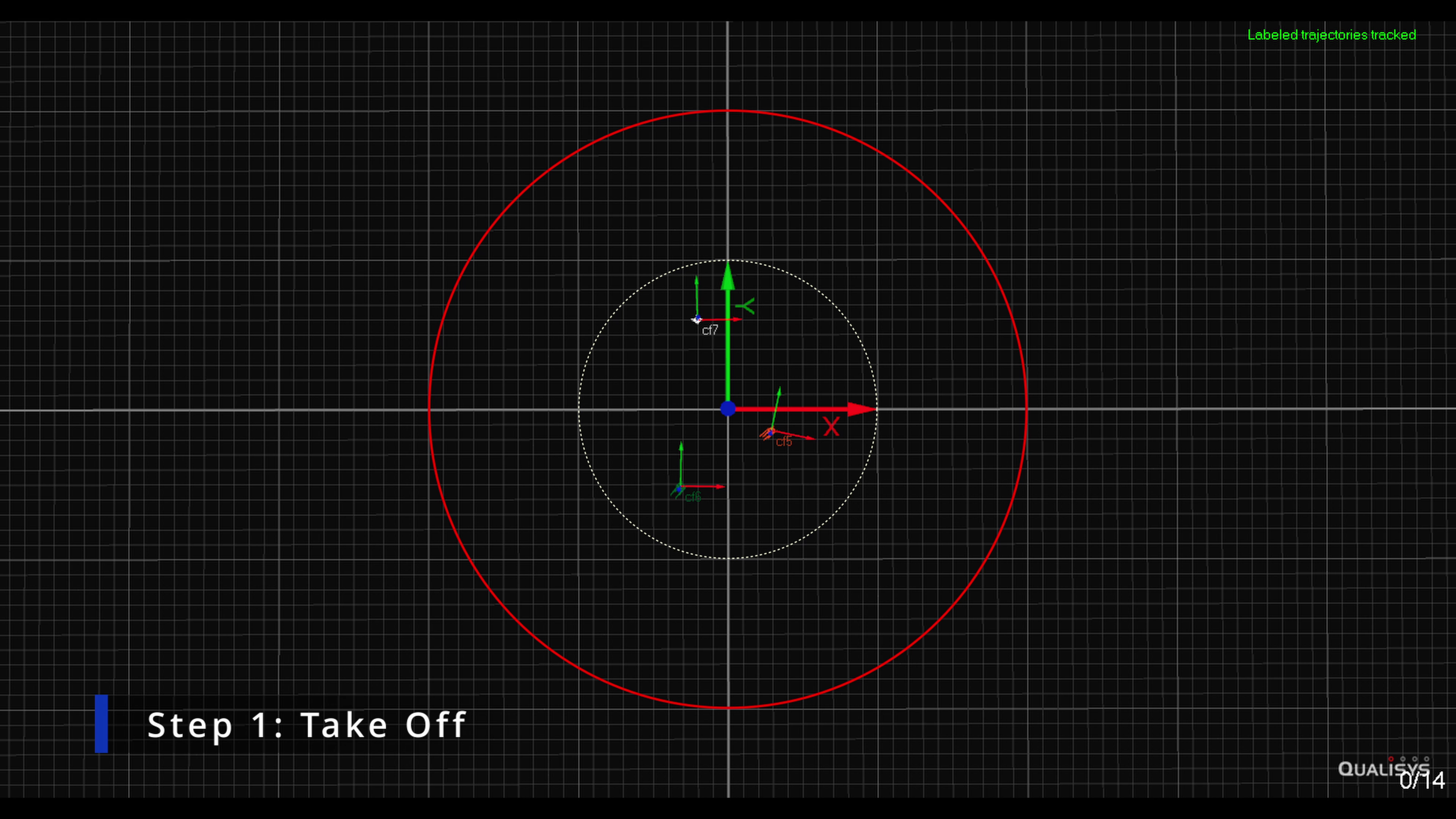}} \hspace*{0.2cm}
 	\subfigure[Moving to actual initial positions]{\includegraphics[scale=0.0483]{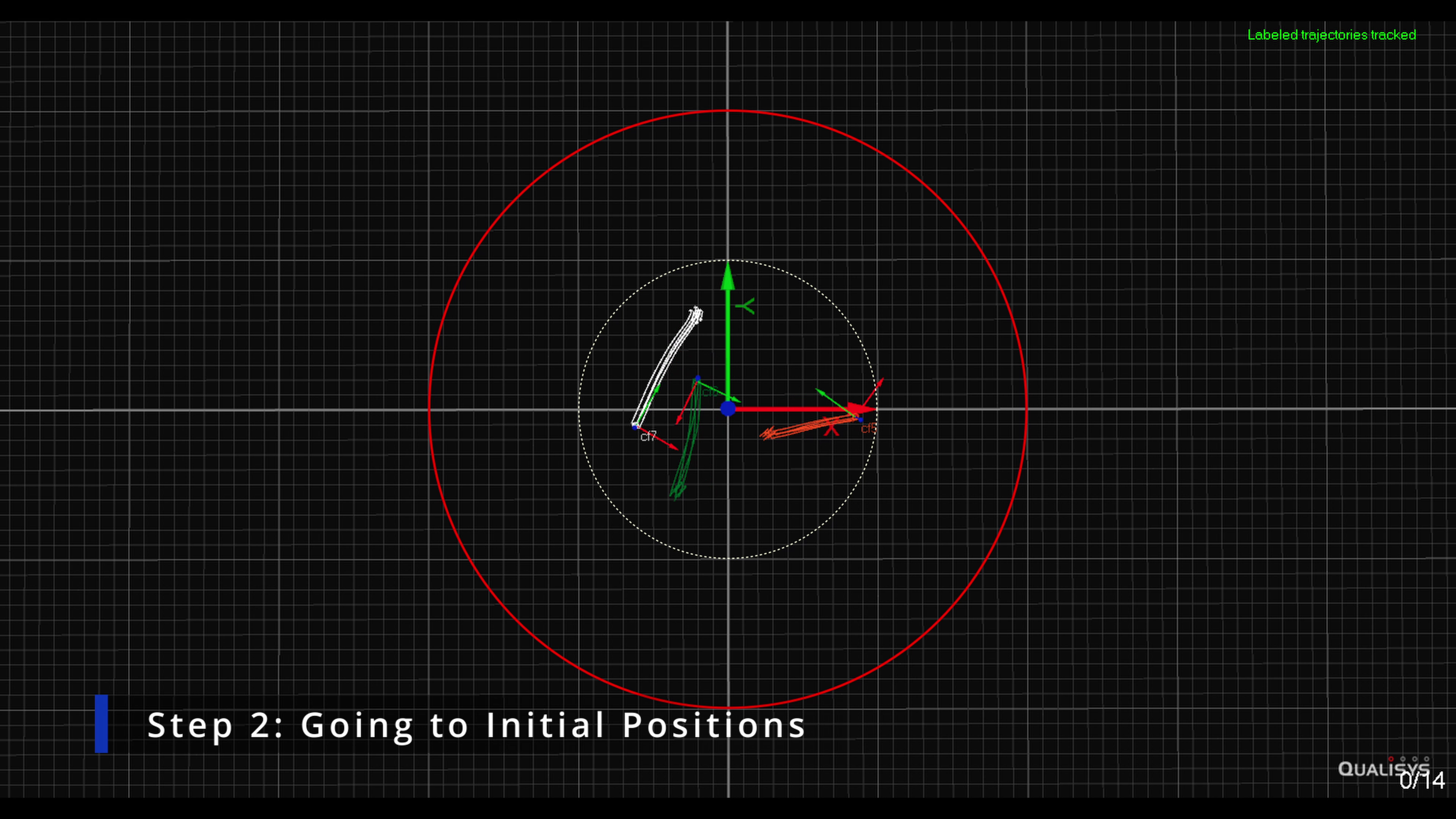}}\\ 
 	\subfigure[Circular maneuver]{\includegraphics[scale=0.0483]{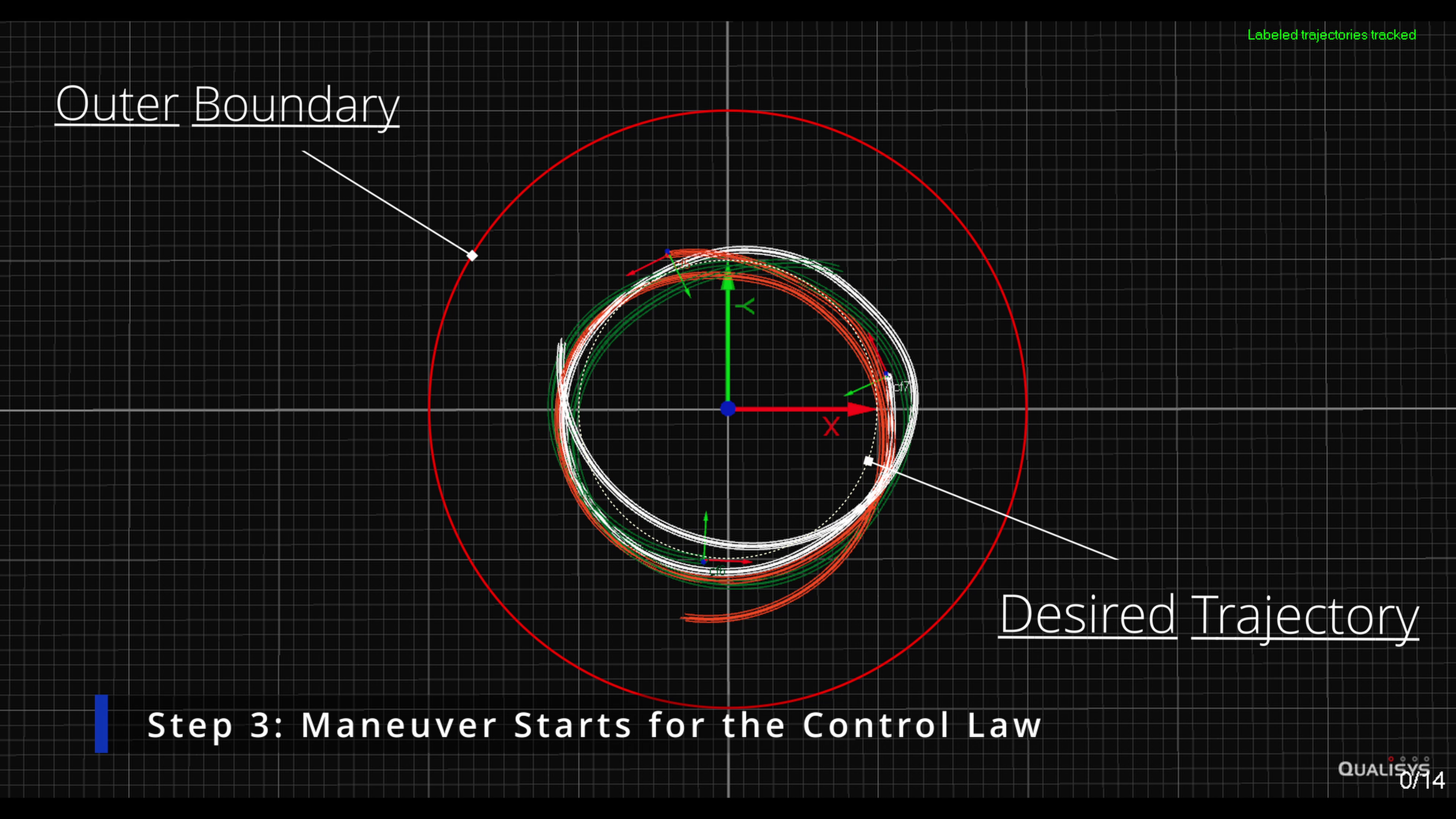}}  \hspace*{0.2cm}
 	\subfigure[Landing after maneuver] {\includegraphics[scale=0.0483]{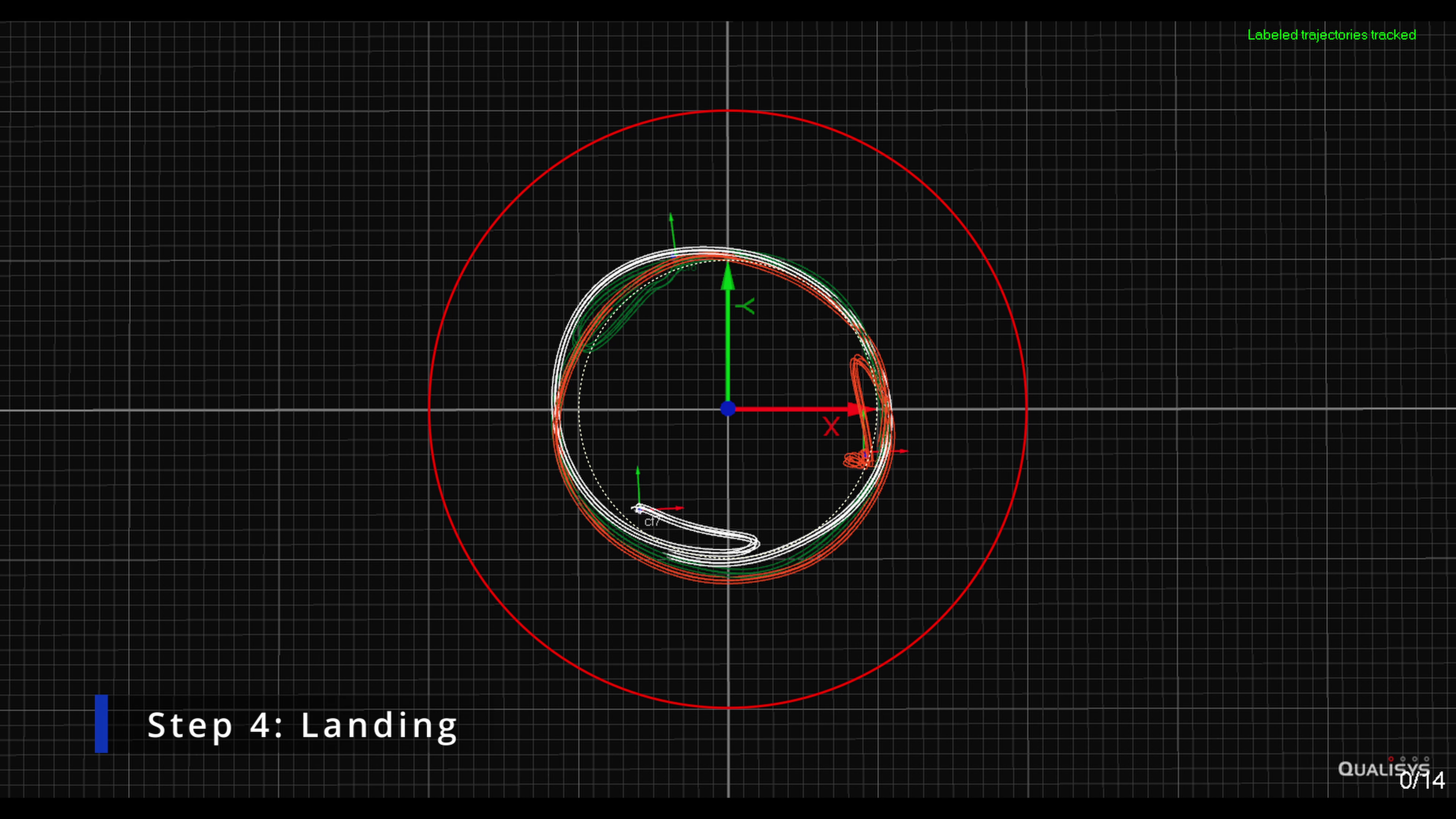}}  
 	\caption{Trajectory-constrained phase-balancing of 3 Crazyflie 2.1 quad-copters around circular curve: Recorded results from QTM software.}
 	\label{circle_steps}
 	\vspace*{-15pt}
 \end{figure*}

\subsection{Simulations}
We simulated the control law \eqref{control} for three agents (i.e., $N = 3$) for both circular and elliptical curves. The center of the curves is taken to be at the origin for both cases whereas the initial conditions of the agents, i.e., $[x_k(0), y_k(0), \theta_k(0)]^T$, are chosen such that the condition $|e_k(0)| < \delta, \ \forall k$ is satisfied for some given $\delta$. As shown in Fig.~\ref{simulations}, the agents move around these curves in phase-balancing (i.e., the angular separation between the neighboring agents is $2\pi/3$ radians), and their trajectories remain bounded within the compact set/boundary marked in red. 

\section{Experimental Setup}\label{section_2}
%The experimental setup comprises three main components: 
\subsection{MoCap System} 
We have a MoCap (Figs.~\ref{mocap1} and \ref{mocap2}) consisting of 10 Qualisys Arqus cameras which cover about a working volume of 75 $m^3$ when calibrated to full usable volume (Fig.~\ref{3dvolume}). Our system is capable of using both passive and active markers for the detection and tracking of rigid bodies.

\subsection{Software Setup}
The software setup has mainly 3 components, as discussed below in detail, and is also shown in Fig. \ref{rqt}. This figure also indicates the flow of data starting from the MoCap cameras to the end goal that are Crazyflie 2.1 quadcopters in our case.

\begin{itemize}
	\item Qualisys Track Manager (QTM): QTM software provides marker distance data, collected by each camera, into readable pose data, including position and orientation. It facilitates the data transmission through a network to the connected systems. We utilize QTM to create three rigid bodies each represented by a unique combination of passive markers on the Crazyflie quadcopters. 
	
	\item ROS: ROS 2, known as Humble Hawksbill, provides tools for communication between system components. It supports real-time systems, essential for our implementations. We use ROS 2 topics to enable continuous data streams, such as sensor data and robot state, to be published by nodes like \emph{cmd\({\_}\)vel} for velocity control and received by nodes for processing.
	
	\item Crazyswarm2: We rely on the Crazyswarm2 package in ROS, extending its functionality to enable high-level velocity control of Crazyflie 2.1 quadcopters. Modifications to the \emph{crazyflie\({\_}\)server} back-end, along with the Python wrapper, allow us to issue velocity commands in the quadcopter's body frame, simulating unicycle model behavior. This expands upon the default Crazyswarm2 capabilities, which offer real-time position control and the desired trajectory following.
\end{itemize}

\begin{figure*}[t]
	\centering
	\subfigure[Take-off from random positions]{\includegraphics[scale=0.0483]{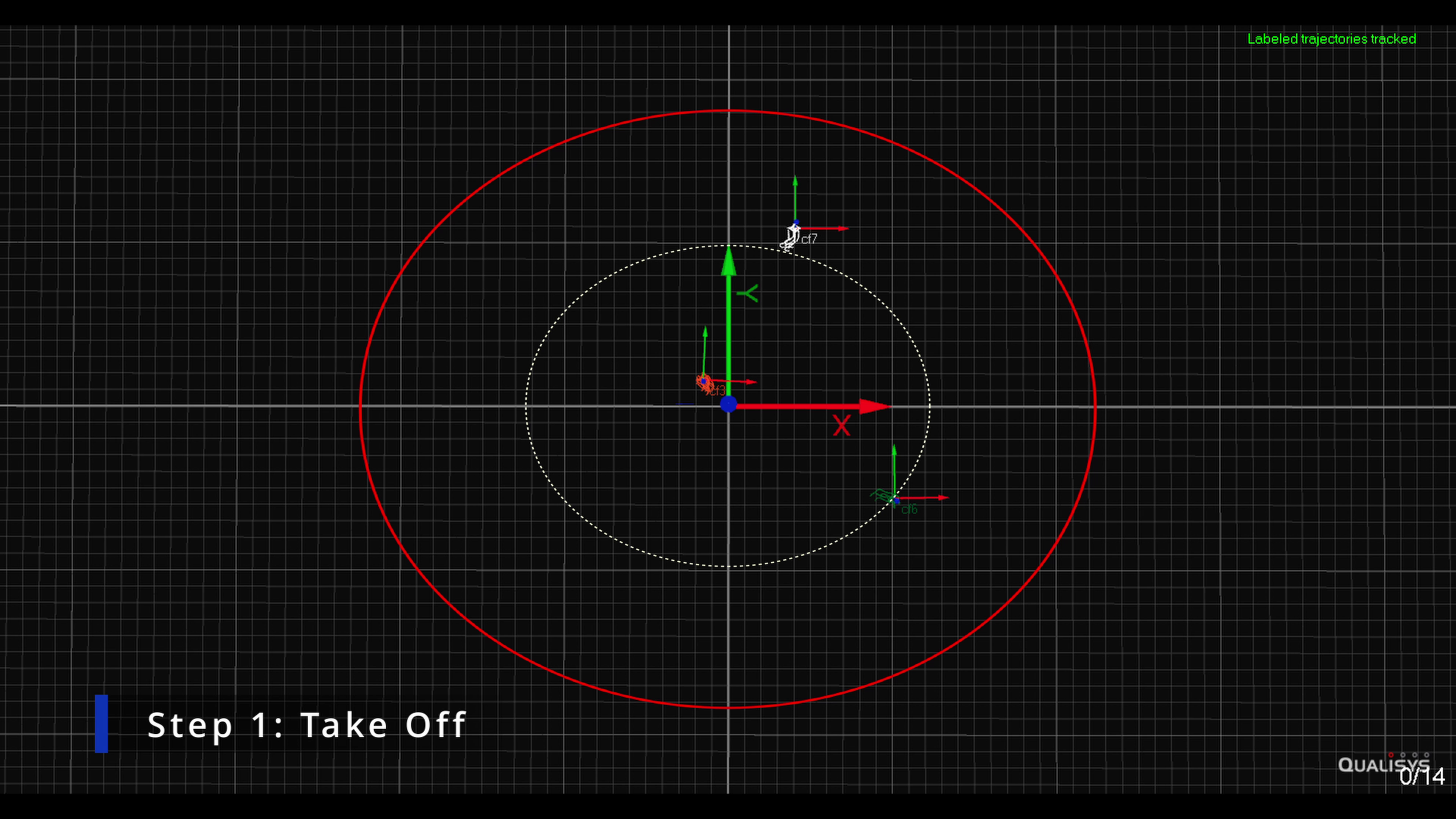}}  \hspace*{0.2cm}
	\subfigure[Moving to actual initial positions]{\includegraphics[scale=0.0483]{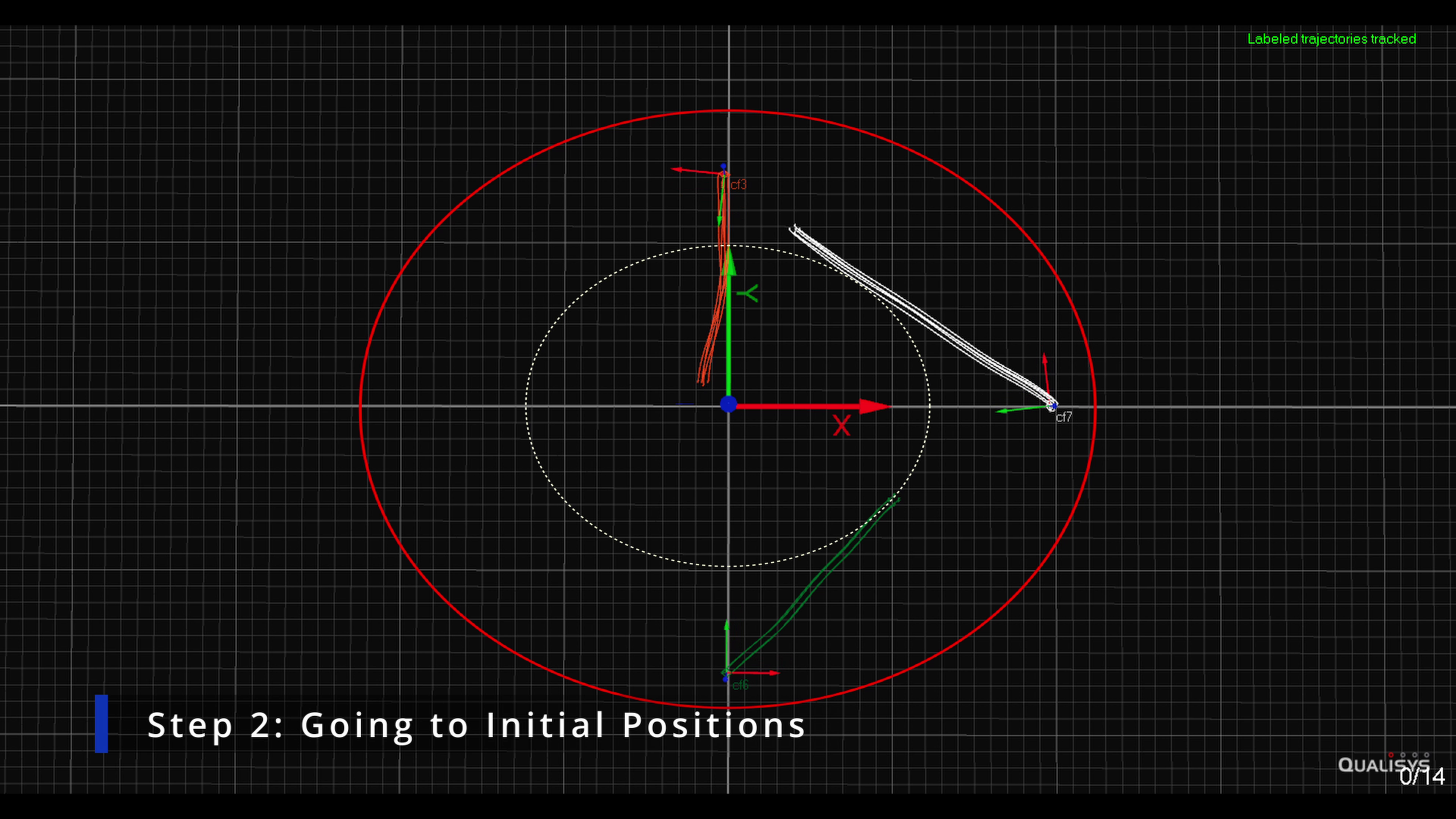}} \\
	\subfigure[Elliptical maneuver]{\includegraphics[scale=0.0483]{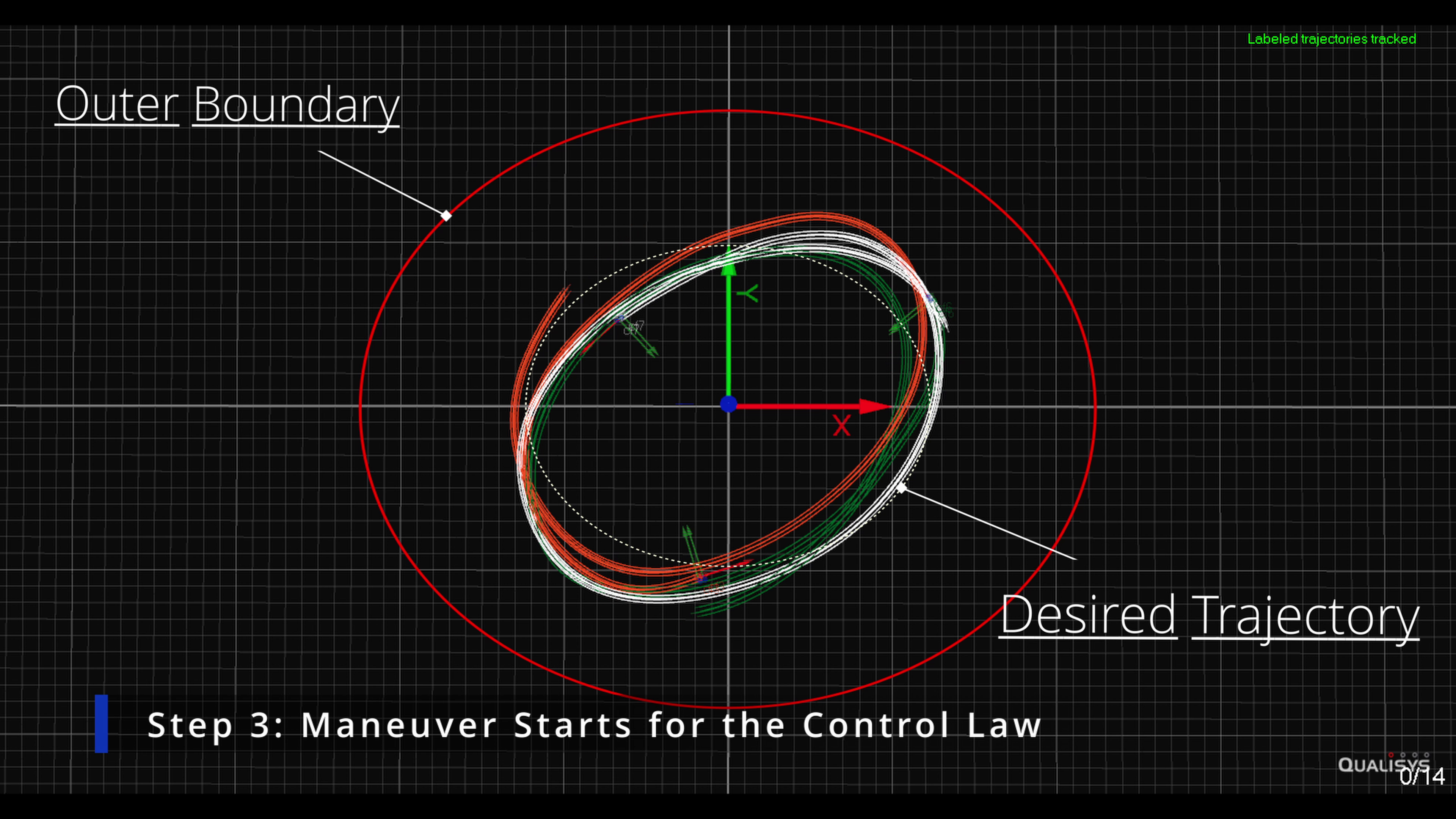}} \hspace*{0.2cm}
	\subfigure[Landing after maneuver]{\includegraphics[scale=0.0483]{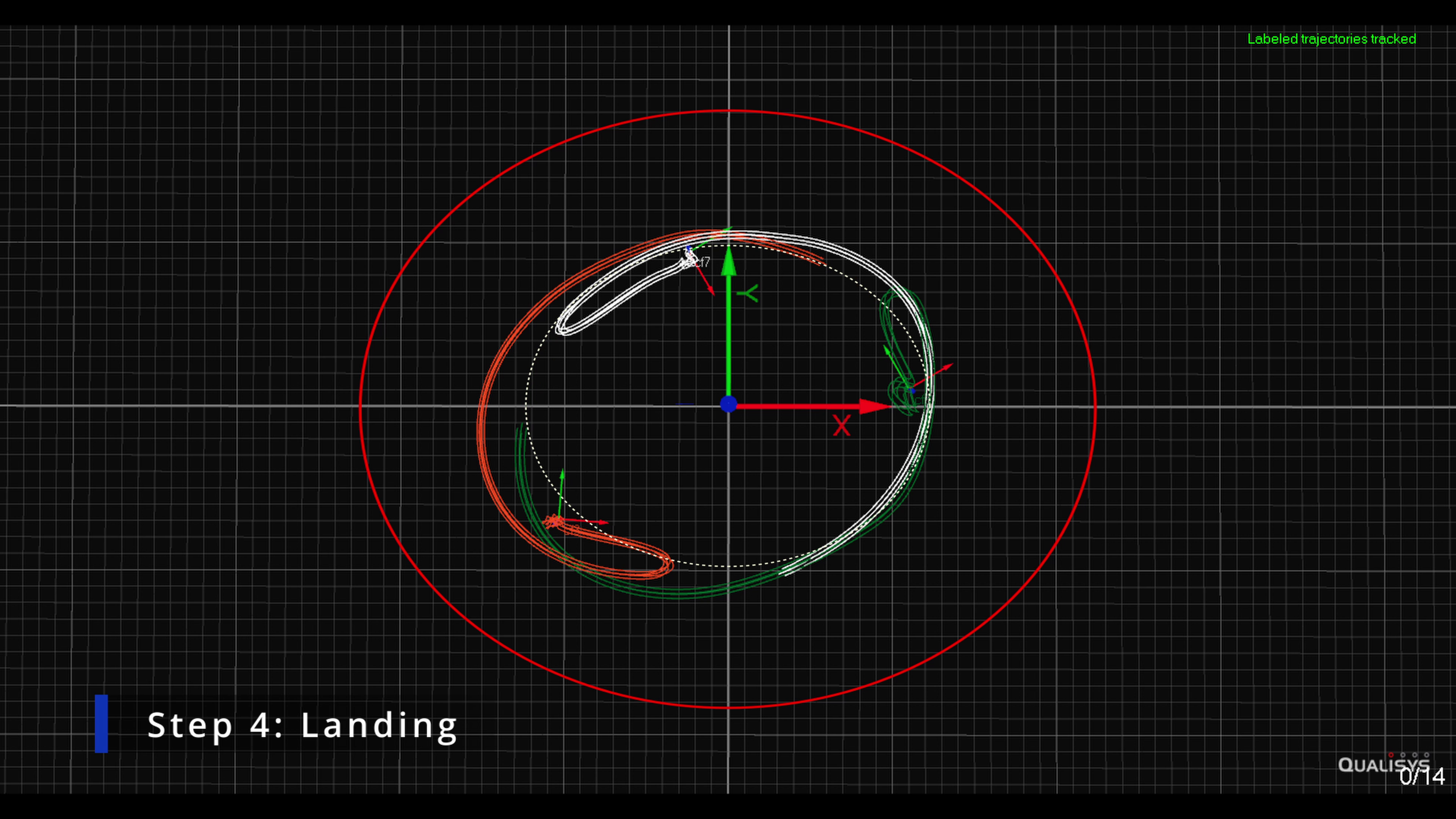}}  
	\caption{Trajectory-constrained phase-balancing of 3 Crazyflie 2.1 quadcopters around elliptical curve: Recorded results from QTM software.}
	\label{ellipse_steps}
	\vspace*{-15pt}
\end{figure*}

 \begin{figure}[t]
	\centering
	\hspace*{-10pt}\subfigure[Circular motion]{\includegraphics[scale=0.33]{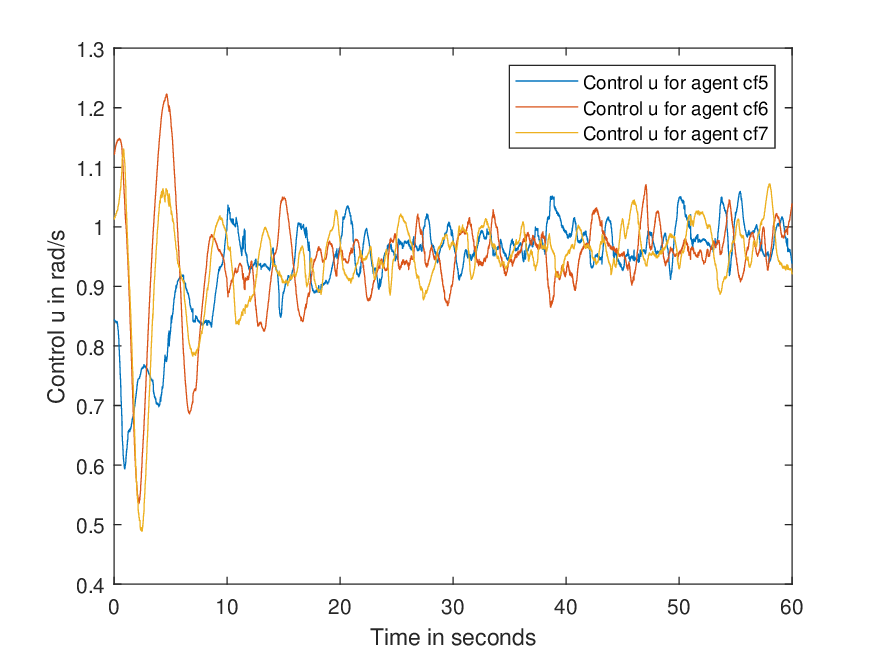}\label{circle_u}}\hspace*{-0.5cm}
	\subfigure[Elliptical motion]{\includegraphics[scale=0.33]{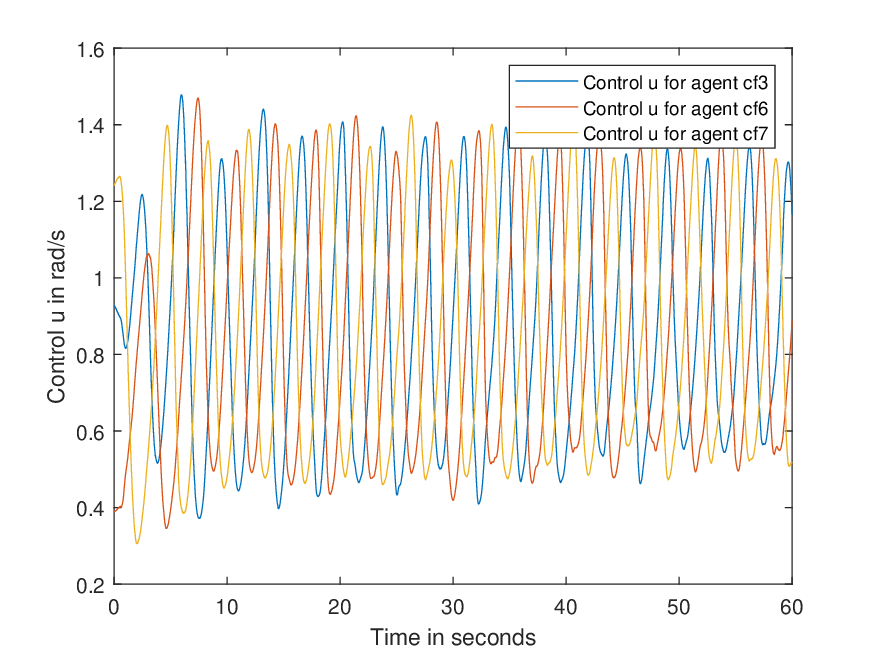}\label{ellipse_u}} 
	\caption{Control laws $u_k$ in case of circular and elliptical motions.}
	\label{control_u}
	\vspace*{-15pt}
\end{figure}

\subsection{Crazyflie 2.1}
We have performed experiments on a swarm of 3 Crazyflie 2.1 quadcopters, as shown in Fig.~\ref{cf_three} where passive markers are fixed using MoCap marker decks. Each unit contains four propellers and weights around 27g. Interested readers can explore further details about it on bitcraze website.

\section{Experimental Results and YouTube Links}\label{section_3}
We conducted two experimental trials employing three Crazyflie 2.1 quad-copters, successfully achieving stable trajectory tracking along the desired curves with bounded motion. In both trials, the quad-copters initiated take-off from random positions, subsequently autonomously realigning themselves to their predefined initial configurations, prior to executing controlled maneuvers. The quad-copters are given unique IDs and unique hexadecimal URIs to differentiate them from each other. 

\subsection{Bounded phase-balancing around a circular curve}
We performed an experiment for the following values of parameters: $r = 1$ m, $\delta = 1$ m, and the initial conditions of the quad-copters are chosen such that $|e_k(0)| < \delta$ for all $k = 1, 2, 3$, and are given by $x(0) = [1.5, 0, -1]^T$, $y(0) = [0, 1.2, -1]^T$ and $\theta(0) = [\pi/2, \pi, -\pi/4]^T$. The snapshots of the recorded experiments in QTM software are shown in Fig.~\ref{circle_steps} at different time instances. Clearly, the quad-copters are at an angular separation of $2\pi/3$ radians and their trajectories are bounded with a circular boundary (shown in \textcolor{red}{red}) of radius $2$ m. Here, the Crazyflie 2.1 quad-copters are identified by the indexing {\bf cf5}, {\bf cf6} and {\bf cf7}. 

\subsection{Bounded phase-balancing around an elliptical curve}
In this case, we have chosen $a = 1.25$ m, $b = 1$ m, $\delta = 1$ m, and the initial conditions of the quadcopters are chosen as $x(0) = [0, 0, 1.75]^T$, $y(0) = [1.2, -1.5, 0]^T$ and $\theta(0) = [\pi,  0, \pi/2]^T$ such that $|e_k(0)| < \delta$ for $k = 1, 2, 3$.  The snapshots of the recorded experiments are shown in Fig.~\ref{ellipse_steps}. Clearly, the quad-copters are at an angular separation of $2\pi/3$ radians and their trajectories are bounded with an elliptical boundary (shown in \textcolor{red}{red}). Here, the Crazyflie 2.1 quadcopters are identified by the indexing {\bf cf3}, {\bf cf6} and {\bf cf7}. 

Further, the control law (or turn rate) $u_k, \forall k$ for both the cases is plotted in Fig.~\ref{control_u}. Fig.~\ref{circle_u} shows that the $u_k \approx 1$ rad/sec, at the steady-state, in case of circular motion. Whereas, $u_k$ varies with time in Fig.~\ref{ellipse_u}, as the curvature $\kappa(\phi)$ is not constant for the elliptical curve. 

%The YouTube video links of both the experiments are provided in the below box:

\begin{table}[h]
\begin{tcolorbox}
\centering{\bf YouTube Video Link} \\ \vspace*{4pt}
\hspace*{-0.4cm}The video demonstration of both of the above cases can be seen on YouTube playlist  on \\ \vspace*{2pt} \hspace*{-0.4cm}\textcolor{magenta}{\url{https://www.youtube.com/playlist?list=PLHjWZ6Jy6e2HCY9EQXf90tEohbSxe8TRn}}\\ \vspace*{4pt}
{\bf Note}: Please copy and paste the link directly in the browser for navigating to the playlist.
\end{tcolorbox}
\vspace*{-15pt}
\end{table}

\section{CHALLENGES AND FUTURE ASPECTS}\label{section_4}
During the course of experiments, we faced several software and hardware-related challenges; the main hurdle was adapting the kinematic model for Crazyflie 2.1 quad-copters as there were no functional velocity control methods in Crazyswarm2 back-end. We did it by modifying the Crazyswarm2 back-end, including \emph{crazyflie\_server}. The other challenge faced by us was regarding the use of incomplete elliptical integral of the second kind which is actually making the calculation lag while the swarm of quad-copters is in motion. This is also reflected in Fig.~\ref{ellipse_steps} where the major and minor axes of the ellipse are rotated by a \emph{small} angle with respect to the origin.

As part of future research, we aim to implement the same on a swarm of ground robots which are more stable and powerful in the sense that the motion of one agent does not affect the motion of another agent, unlike small aerial robots like Crazyflie 2.1. It would be interesting to explore real-world scenarios without using the MoCap system and overcome the challenges it brings when we move to an outdoor environment.

%%%%%%%%%%%%%%%%%%%%%%%%%%%%%%%%%%%%%%%%%%%%%%%%%%%%%%%%%%%%%%%%%%%%%%%%%%%%%%%%%

\bibliographystyle{IEEEtran}
\bibliography{References}

\end{document}